# Extended Range Telepresence for Evacuation Training in Pedestrian Simulations

A. Pérez Arias[1] and U. D. Hanebeck[1], P. Ehrhardt[2], S. Hengst[2], T. Kretz[2], and P. Vortisch[2]

[1]Intelligent Sensor-Actuator-Systems Laboratory (ISAS)
Institute for Anthropomatics, KIT, Germany

[2]PTV Planung Transport Verkehr AG
 Stumpfstraße 1, D-76131 Karlsruhe, Germany

Corresponding author: antonia.perez@kit.edu

**Abstract:** In this contribution, we propose a new framework to evaluate pedestrian simulations by using Extended Range Telepresence. Telepresence is used as a virtual reality walking simulator, which provides the user with a realistic impression of being present and walking in a virtual environment that is much larger than the real physical environment, in which the user actually walks. The validation of the simulation is performed by comparing motion data of the telepresent user with simulated data at some points of the simulation. The use of haptic feedback from the simulation makes the framework suitable for training in emergency situations.

## Introduction

Telepresence allows visiting remote or virtual places with a high degree of realism and offers new tools for human motion understanding. Applications of Telepresence include teleoperation, military training, driving and flying simulators, etc.

The feeling of presence is achieved by visual, acoustic, and haptic sensory information recorded from the remote environment and presented to the user on an immersive display. The more the user's senses are involved, the better the immersion in the target environment is. In order to make use of the sense of motion, which is especially important for human navigation and way-finding, the user's motion is tracked and transferred to the avatar in the target environment. This is known as Extended Range Telepresence, and enables the user to make use of proprioception, the sense of motion, to navigate the avatar intuitively by natural walking, instead of using devices like joysticks, keyboards, pedals, or steering wheels. To allow exploration of an arbitrarily large target environment while moving in a limited user environment, we developed Motion Compression [1]. By preserving the walked distances and the turning angles, Motion Compression transforms the path in the target environment into a feasible path in the user environment, and guides the user on his path. Fig. 1 shows the user interface in the Extended Range Telepresence system.

Through the egocentric view, this framework provides a first person evaluation of the simulation. The user is not passively looking at the simulation, but he feels present in the simulation and can interact with other pedestrians. Extended Range Telepresence also allows the evaluation of pedestrian models based on gathered real user data. These experiments in the virtual environment are not only cheap to set up, but are also quick to evaluate, as all position data of the user is available anyway.

An application of our framework with particular focus on gaining spatial knowledge is the training of evacuations, where people are trained to find the way out of buildings, ships, or planes, so that planners could check how intuitively occupants find their way out before beginning with the real-world construction.

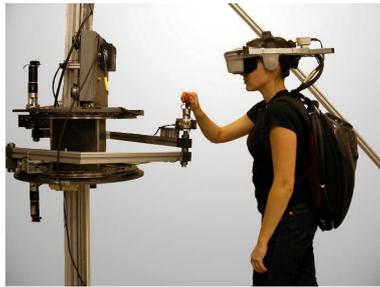

**Fig. 1: User interface in the Extended Range Telepresence system.**

This work aims at employing Extended Range Telepresence first as a tool for evaluating and calibrating pedestrian simulations and second to train people in emergency evacuations using the validated simulations. In order to increase the degree of immersion and to present additional information to the user, the use of haptic information is also explored. Finally, a route choice scenario is examined and the potentials of the system in various fields of application are discussed.

# Extended Range Telepresence with Motion Compression

In order to allow for exploration of an arbitrarily large target environment while moving in a limited user environment, Motion Compression provides a nonlinear transformation between the desired path in the target environment, the target path, and the user path in the user environment. The algorithm consists of three functional modules (see Fig. 2).

First, the *path prediction* gives a prediction of the desired target path based on the user's head motion and on knowledge of the target environment. If no knowledge of the target environment is available, the path prediction is based completely on the user's view direction. Second, the *path transformation* transforms the target path into the user path in such a way, that it fits into the user environment. In order to guarantee a high degree of immersion, the user path has the same length and the same turning angles as the target path. The two paths differ, however, in path curvature. The nonlinear transformation found by the path transformation module is optimal regarding the difference of path curvature. Fig. 3 shows an example of the corresponding paths in both environments. Finally, the *user guidance* steers the user on the user path, while he has the impression of actually walking along the target path. It benefits from the fact that a human user walking in a goal oriented way constantly checks for his orientation toward the goal and compensates for deviations. By introducing small deviations in the avatar's posture, the user can be guided on the user path. More details can be found in [1, 2].

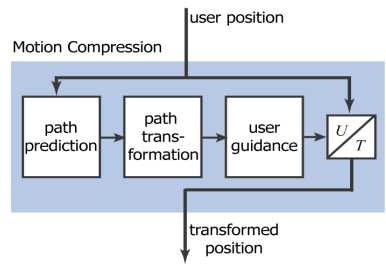

**Fig. 2: Motion Compression.**

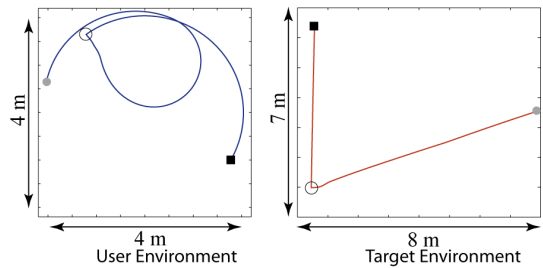

**Fig. 3: Path in both environments.**

## Evaluation of the Pedestrian Simulation

A common problem in pedestrian simulations is the lack of available data for the calibration of pedestrian models [10]. Not all kinds of real experiments can be replaced by virtual experiments – for example high density situations with many contacts would be difficult – but Extended Range Telepresence should provide a suitable framework for gathering data, for example, for route choice issues and obviously in scenarios that involve some risk for the test participants. Furthermore, Extended Range Telepresence represents a systematic evaluation tool, since all data is logged, while trajectory extraction of real experiments takes quite some effort [11-13].

First, it is necessary to specify reproducible test set-ups in the proposed framework. To do this, it is required to compare data from real experiments with data from experiments in the virtual reality, in order to identify the situations in which the behavior of the user in the Extended Range Telepresence system differs most from real situations. We expect that the user's behavior will be more realistic if the quality of the virtual scenario increases. For this reason, we intend to investigate, how haptic information and a realistic representation of the pedestrians' morphology affect the user's behavior and his motion parameters (e.g., position, velocity, distance to next pedestrian, etc.). This will allow us the definition of reasonable criteria to assess test results.

The evaluation of the simulation consists in augmenting the simulation with a real user. The user's state (position, orientation, and velocity) is used to initialize the simulation, and at every time step the state of the telepresent user and the simulated state are compared. An optimization is then performed in order to find the parameters that minimize the deviation between actual and simulated state. Global parameters of the simulation that affect the route choice and the acceptance of the simulated traffic flow will also be evaluated.

Finally, owing to the fact that unexpected and dangerous escape situations exclude real-life experiments [7,14], it is desired to find suitable parameters on escape panics, which permit to use the pedestrian simulation also in evacuation scenarios.

## Evaluation Framework

A connection of Extended Range Telepresence and the VISSIM [3] pedestrian simulation was implemented [4] so that the scene shown to the user is populated with pedestrians from a VISSIM simulation.

The framework was extended in order to introduce haptic information into the simulation, since it is indispensable for increasing the degree of immersion and to present additional information to the user, e.g., by displaying the contact force against other pedestrians or by signalizing a forbidden or too dangerous path, which must be avoided. A haptic interface especially designed for Extended Range Telepresence provides haptic information from the target environment [5]. The data flow in the framework is shown in Fig. 4.

The user's posture is tracked and fed into the system. Every time an update of the user's position is available, the Motion Compression algorithm is executed, and the transformed avatar's posture is calculated and sent to the simulation. The simulation constantly captures live images from the avatar's view, which are sent to the user and presented on a head-mounted display, so that the user has the impression of walking, without any restriction, in the simulation.

The simulation also calculates the resulting force acting on the user, which results from the contact with obstacles and other pedestrians, and sends this to the Motion Compression server, which transforms the force from the target to the user environment. The resulting force vector has the same magnitude and relative direction to the user path in both environments. The transformed force vector is sent to the haptic interface that displays the force on the user's hand.

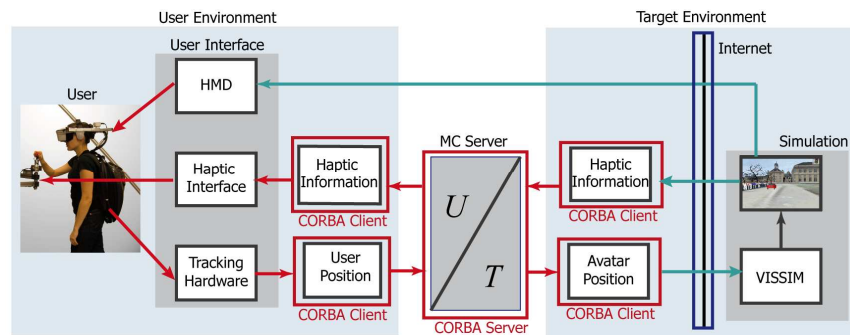

**Fig. 4: Data flow in the proposed framework.**

## Use of Haptic Information

The use of haptic information is indispensable for increasing the sense of presence so that the user behaves realistically when immersed in the simulation. When the simulation is augmented by a real user, the simulated pedestrians react to him and try to avoid him. However, without physical forces displayed on the haptic inter-face, it would be difficult for the user to go ahead into the crowd without walking through simulated pedestrians.

The physical force acting on the user is due to the contact with other pedestrians and with borders of buildings, walls, obstacles, etc. This force can be obtained by computing the repulsive effects with other pedestrians and obstacles when a collision takes place in the direction of motion. The pedestrians touch each other if their distance is smaller than the sum of their radii. In this case, the interaction force is the sum of two forces [7]: a normal force counteracting body's compression and a sliding friction force impeding relative tangential motion. The interaction force with the walls is treated analogously.

Since the Social Force Model [6-9], which models the behavior of a pedestrian in a crowd, does not draw a clear distinction between physical and psychological forces, we apply on the real user, in a first approach, the resultant force calculated by the Social Force Model. As the user sees the simulated pedestrians and the walls in the display (and as the forces in the model are mediated by the senses and the psyche), by including all forces in the resultant force displayed by the haptic interface, the psychological forces are in a way doubled. However, since the physical forces are much higher than the psychological forces during contact, the resultant social force can be regarded as an approximation of the interaction force, just when a collision takes place. Otherwise the applied force is zero.

## Evaluation Scenario

The implemented system has been extensively tested in a number of virtual scenarios (Fig. 5). The setup uses a high-definition head-mounted display of 1280x1024 pixels per eye and a field of view of 60°. The user's posture, i.e., position and orientation, is estimated by an acoustic tracking system that provides 50 estimates per second [15].

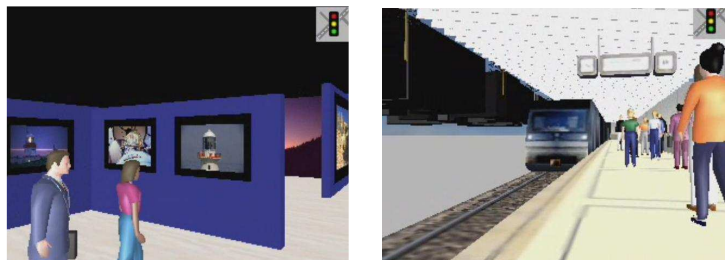

**Fig. 5: Impression of tested scenarios.**

A simple route choice scenario has been selected to illustrate the proposed approach (Fig. 6). Such a scenario is suitable to calibrate VISSIM's simulation module that allocates the pedestrians to the gates. A rigorous calibration of the simulation would require the comparison of virtual with real experiments, as explained in the previous section but we assume here that in such a scenario the user in the Telepresence system behaves realistically.

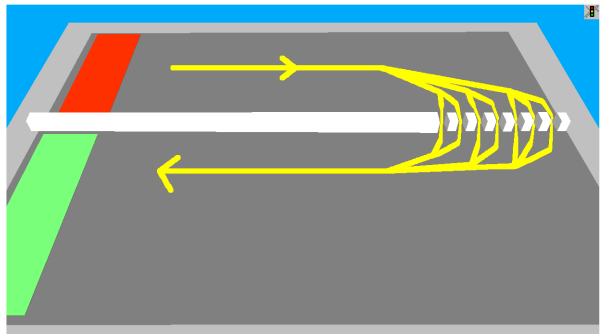

**Fig. 6: Schematic view of route choice scenario used to evaluate VISSIM.**

In this scenario, pedestrians start walking on the red surface, walk across one gate, and disappear on the green surface. In each trial, 150 pedestrians were simulated. The task completion time, the covered distance, and the chosen gate were recorded and compared with those of the test participants. Fig. 7 shows the trajectories of pedestrians and participants.

The framework allows performing reproducible tests that can easily be evaluated. The simulation first calculates a realistic distribution of pedestrians over the gates. Most of pedestrians take the closest gate, as expected, and the others take farther gates in decreasing number. By monitoring together the behavior of real and simulated pedestrians, it is possible to compare their distributions.

It is notable, that most pedestrians in the simulation choose the first gate, whereas three of five test participants take the second one. In the last path segment, the trajectories of pedestrians and participants differ considerably. This is due to the fact that pedestrians in the simulation do not use the entire available surface, but this is not relevant for the experiment. The observation that a real user considers not only the distance but also the availability of the infrastructure and the waiting time in order to choose a farther gate is also in agreement with real observations.

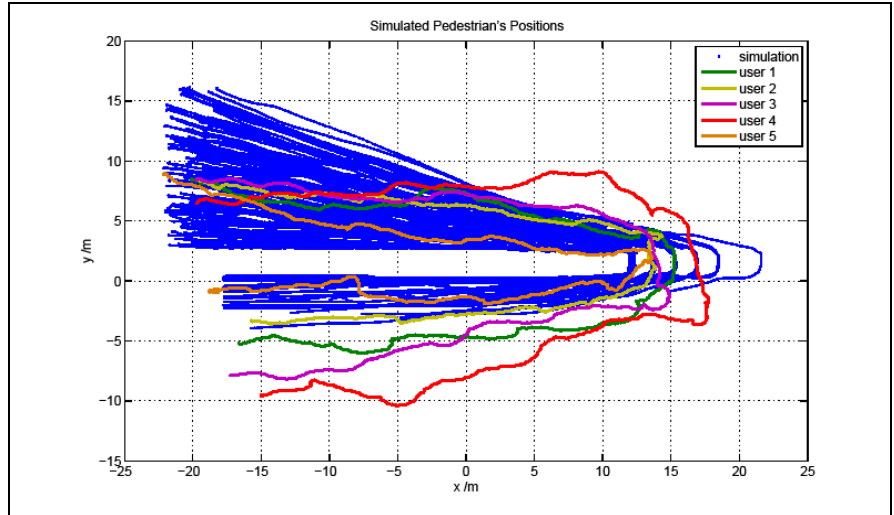

**Fig. 7: Trajectories of simulated pedestrians and test participants.**

The calibration process must be iterative. Test participants behave at the beginning differently from the simulated pedestrians. Once the simulation has been so adjusted that the distribution of simulated pedestrian matches the distribution of test participants, another calibration run must be performed. This procedure, which is known as *dynamic assignment*, must be repeated until it converges. In traffic assignment, there is one iteration method, in which the convergence is demonstrated. It consists in regarding at the iteration *N+1* the past *N* iterations with a weighting factor *1/N*. However, this method converges very slowly and needs a lot of experiments. Therefore, a simplified method is used, which just considers the last iteration (iteration *N*) with a weighting factor *W* and the iteration before (iteration *N-1*) with a weighing factor *1-W*.

## Conclusions and Future Work

Extended Range Telepresence offers an appropriate tool for the evaluation and the calibration of pedestrian simulations, by having one real person walking through a crowd of simulated agents. The main benefit of the proposed framework is that experiments with a real person walking in a virtual environment are cheap and easy to evaluate, since the position of the test person is available at any time during the experiment. Haptic information is used to enhance the realism of the simulation and therefore, the validity of the given evidence.

Currently, it seems very difficult to give a highly immersive and realistic impression for high-density situations, when contacts to other pedestrians occur frequently and from all sides. But the system might prove very useful for virtual experiments on topics where currently only few data are available: route choice [16-19], behavior as participants of road traffic (in VISSIM, vehicles can easily be included to virtual experiments) [20], and behavior in certain dangerous situations [21, 22], which anyway are not directly and fully realistically accessible by experiment. Further work must be done in defining an evaluation metric to prove that the experiments in the Telepresence system are as valid as real experiments for the purpose of improving pedestrian simulations in specific situations, especially in evacuation scenarios.